\input{style/arxiv-general.cfg}
\documentclass[aoas,MSNbibl,nameyear,seceqn,secthm,dvips]{arximspdf}
\makeatletter
   \@ifpackageloaded{graphicx}{}{\usepackage{graphicx}}
\makeatother
\usepackage{stfloats}
\usepackage{graphicx}

\doi{10.1214/14-AOAS777} 
\volume{9}
\issue{1}
\pubyear{2015}
\firstpage{225}
\lastpage{246}
\docsubty{FLA}

\makeatletter
  \fnbelowfloat
\makeatother

\begin{document}
\begin{frontmatter}

\title{A multi-functional analyzer uses parameter constraints to improve the
 efficiency of model-based gene-set analysis}
\runtitle{A multi-functional analyzer}

\begin{aug}
\author[A]{\fnms{Zhishi}~\snm{Wang}\thanksref{m1}\ead[label=e1]{wangz@stat.wisc.edu}},
\author[A]{\fnms{Qiuling}~\snm{He}\thanksref{m2}\ead[label=e2]{ally00ling@gmail.com}},
\author[B]{\fnms{Bret}~\snm{Larget}\thanksref{m3}\ead[label=e3]{larget@stat.wisc.edu}}
\and\\
\author[C]{\fnms{Michael A.}~\snm{Newton}\corref{}\thanksref{m4}\ead[label=e4]{newton@stat.wisc.edu}\ead[label=u1,url]{http://www.stat.wisc.edu/\textasciitilde newton/}}
\runauthor{Wang, He, Larget and Newton}
\affiliation{University of Wisconsin, Madison}
\address[A]{Z. Wang\\
        Q. He \\
        Department of Statistics\\
        University of Wisconsin, Madison \\
     1300 University Avenue\\
        Madison, Wisconsin 53706\\
        USA\\
 \printead{e1} \\
\phantom{E-mail:\ }\printead*{e2}}
\address[B]{B. Larget \\
        Departments of Statistics and Botany \\
        University of Wisconsin, Madison \\
       1300 University Avenue \\
 Madison, Wisconsin 53706 \\
 USA\\
\printead{e3}}
\address[C]{M. A. Newton \\
 Departments of Statistics and Biostatistics\\
 \quad  and Medical Informatics\\
 University of Wisconsin, Madison \\
1300 University Avenue \\
 Madison, Wisconsin 53706\\
 USA\\
\printead{e4}\\
\printead{u1}}
\end{aug}
\thankstext{m1}{Supported in part by a fellowship from the Morgridge Institute of Research.}
\thankstext{m2}{Supported in part by a research assistantship from the National Institutes of Health (NIH) (R21 HG006568);
        currently employed by Novartis Pharmaceuticals.}
\thankstext{m3}{Supported in part by NIH R01 GM086887 and NSF DEB 0949121.}
\thankstext{m4}{Supported in part by NIH R21 HG006568.}

\received{\smonth{10} \syear{2013}}
\revised{\smonth{8} \syear{2014}}

\begin{abstract}
 We develop a model-based methodology for integrating gene-set information
with an experimentally-derived gene list.  The methodology uses a
previously reported sampling model, but takes
advantage of natural constraints in the high-dimensional discrete parameter
space in order to work from a more structured prior distribution than
is currently available.  We show
how the natural constraints are expressed in terms of linear inequality
constraints within a set of binary latent variables.  Further, the currently available
 prior gives low probability to these constraints in complex systems, such as Gene Ontology (GO),
 thus reducing the efficiency of statistical inference.
We develop two computational advances to enable posterior inference
within the constrained parameter space: one using integer linear programming
for optimization and one using a penalized Markov chain sampler.
Numerical experiments demonstrate the utility of the
new methodology for a multivariate integration of genomic data with
GO or related information systems. Compared to available methods, the proposed
multi-functional analyzer covers more reported genes without mis-covering nonreported
genes, as demonstrated on genome-wide data from  association studies of type~2
diabetes and from RNA interference studies of influenza.
\end{abstract}

\end{frontmatter}

\section{Introduction}

In statistical genomics, the gene list is a recurring data structure.  We have in mind
 situations where experimental results
amount to a collection of genes measured to have some property.  Examples
include the following: RNA expression studies, in which the property might be differential
expression of the gene between two cell types; genome-wide RNA knock-down
studies, in which the property is significant phenotypic alteration caused
by RNA interference; chromatin studies
recording genes in the vicinity of transcription factor binding sites or having
certain epigenetic marks.
 In all cases, the reported gene list is really the result
of inference from more basic experimental data.  These more basic data may be available
to support subsequent analyses, but we are concerned with the important and
 relatively common case in which the gene list itself is the primary data set brought
 forward for analysis.

\setcounter{footnote}{4}

 The statistical question of central importance in the present paper is how to interpret
 the gene list in the context of preexisting biological knowledge about
 the functional properties of all genes, as these exogenous data are recorded in database
 systems, notably Gene Ontology (GO), the Kyoto Encyclopedia (KEGG) and the reactome,
 among others [\citet{go}; \citet{kegg}; \citet{m08};
        \citet{bioc}].
  For us, exogenous data form a collection of gene sets, with each set
 equaling those genes previously determined, by some evidence, to have a specific
 biological property.  Recently, for example, the full GO collection contained 16{,}527
 sets (GO terms) annotating 17{,}959 human genes.\footnote{Bioconductor package org.Hs.eg.db, version 2.8.0.}
 Needless to say, genes
 are typically annotated to multiple gene sets (median 7 sets per gene among the genes
  annotated to sets which contain between 3 and 30 genes, e.g.), covering all sorts of
 functional properties.  The task of \textit{gene-set analysis} is to efficiently interpret
 the functional content of an experimentally-derived gene list by somehow integrating
 these endogenous and exogenous data sources [\citet{k12}].

 Our starting point is an exciting development in the methodology of gene-set
 analysis.  Model-based gene-set analysis (MGSA) expresses gene-level indicators of
 presence on the gene list as Bernoulli trials whose
 success probabilities are determined in a simple way by latent activity states of
 binary variables associated with the gene sets [\citet{b10};
\citet{b11}].
 Inference seeks
 to identify the \textit{active} gene sets, as these represent functional drivers of
 the experimental data.  Inference is computationally difficult because the activity
 state of a given gene set depends not only on experimental data for genes in
that set, but also on the unknown
 activity states of all other gene sets that annotate
 these same genes.
 MGSA overcomes the problem through Bayesian inference implemented
 with an efficient Markov chain Monte Carlo (MCMC) sampler, and thus provides
 marginal posterior probabilities that each gene set is in the active state.  The
 MGSA  methodology is compelling. Because it treats all gene sets in the collection
 simultaneously, it provides a truly multivariate analysis of the exogenous data source,
 where most available approaches are univariate (one set at a time).
 Where set/set overlaps are a nuisance in most gene-set methodologies, MGSA utilizes
 them directly in modeling and inference.  This accounts for pleitropy, that
 genes have multiple biological functions, reduces the risk of
 spurious associations, and leads to cleaner output whereby a typical list of gene sets
inferred to be active is simpler and exhibits less redundancy than in
 standard univariate analyses [\citet{b10}; \citet{n12}].

Our analysis reveals a feature of MGSA that adversely affects its
 statistical properties. In ever denser
collections of gene sets, the MGSA prior distribution puts more and more mass on
logically inconsistent joint activity states.  As a result, data need to work
 ever harder to overcome this misguided prior probability.  The
effect is tangible;  for a given amount of data, fewer truly activated
 gene sets are inferred to be active, compared to what is achievable with an
alternative formulation.
We propose a new methodology, the multi-functional analyzer (MFA), which aims to
improve the statistical efficiency of MGSA. It uses two computational advances
that enable posterior inference in the high-dimensional constrained space of
 joint activity states. One is an efficient MCMC sampling scheme constructed
 by penalizing the log-posterior in the unconstrained space
 and one is a discrete optimizaton scheme that
translates the inference problem into an integer linear programming (ILP) problem.

We note that  inference about gene-set
 activity states may be interesting from the general perspective
 of high-dimensional statistics.  Typically, dependence among data from
different inference units (sets, in this case) is considered a nuisance and
 testing aims to identify nonnull units (active sets) by some methodology that is
 robust to dependencies, since these dependencies are often difficult to
  estimate from available data.  In the present
 context dependencies are complicated but explicit, and inference benefits
 by using them to advantage. Finally, we also note that the probability model underlying
 our methodology---the \textit{role model}---has potential utility
 in other domains of application. It provides a simple way to relate data collected
 at one level (genes, in this case) to inference units that are unordered collections
 of the former (gene sets, in this case).

\section{Role model}

\subsection{Model}

Following the description in Newton et al., we have a finite number
 of \textit{parts} $p$ and a finite number of \textit{wholes} $w$, where each whole is an
unordered set of parts.  The incidence matrix $I= (I_{p,w})$ is determined
from external knowledge, where $I_{p,w}=1$ if and only if $p \in w$.  The intended correspondence
is that genes are parts and gene sets (i.e.,
functional categories) are wholes.
The matrix $I$ encodes a full collection of gene sets.  We will have measured data on
the parts and aim to make inference on properties of the wholes.

The experimentally-derived gene list may be viewed as a vector of Bernoulli trials
$X=(X_p)$, with $X_p=1$ if and only if part (gene) $p$ is on the list.  First proposed
in \citet{b10}, the role model describes the joint distribution of $X$
in terms of latent binary $(0/1)$ activity variables $Z= (Z_w)$ and by
part-level activities induced by them:
$A_p = 1$ if   $Z_w = 1$ for any $w$ with $p \in w$ or, equivalently,
\begin{equation}
\label{eqf1}
A_p = \max_{w\dvtx  p\in w} Z_w.
\end{equation}
This conveys the simple assumption that a part is activated if it is in any whole
that is activated.  For false-positive and true-positive parameters $\alpha, \gamma \in (0,1)$,
with $\alpha < \gamma$, the
model for $X$ entails mutually independent components (conditionally on latent
activities), with
\begin{equation}
\label{eqmodel}
X_p \sim \operatorname{Bernoulli} \cases{
 \alpha, &\quad\mbox{if $A_p=0$},
\cr
\gamma, &\quad \mbox{if $A_p=1$}.}
\end{equation}
Simply, activated parts (i.e., those with $A_p=1$) are delivered to the list
at a higher rate than inactivated parts.   A key feature of the model is that a part
(gene) is activated by virtue of any one of its functional \textit{roles}; this implies
that a gene may be activated and yet be part of a functional category that is
 inactivated, which  is in contrast to most other gene-set inference methods
 [e.g., \citet{gb}; \citet{bnr}; \citet{sa09}] and
 which provides for a fully multivariate analysis of the gene list.
In \citet{b10} it is
further assumed, for the sake of Bayesian analysis, that uncertainty in whole-level
activities is represented with a single rate parameter $\pi \in (0,1)$:
\begin{equation}
\label{eqprior1}
Z_w \sim_{\mathrm{i.i.d.}} \operatorname{Bernoulli} (\pi).
\end{equation}
Taken together, the model (\ref{eqmodel}) and the prior (\ref{eqprior1}) determine
a joint posterior for~$Z$ given $X$.
The \texttt{R} package \texttt{MGSA} (model-based gene-set analysis)
reports MCMC-computed marginal posterior probabilities $P(Z_w=1|X)$,
 also integrating uncertainty in the system parameters $\alpha$, $\gamma$
and $\pi$, and thus provides a useful ranking of the wholes
 [\citet{b11}; \citet{rdct}].

In addition to the system incidence matrix $I$, a useful data structure for computations
turns out to be the bipartite graph $\mathcal{G}$, having whole nodes and part nodes,
and an edge between $w$ and $p$ if and only if $I_{p,w} = 1$ (i.e., iff $p \in w$).

\subsection{Activation hypothesis}

As defined above, the role model allows that a~whole can be inactive while all of its
parts are active.  This can happen because of overlap among the
wholes. Specifically, if $w$ is contained in the union of other wholes $\{w'\}$, then
all $Z_{w'}=1$ will force $A_p=1$ for all $p \in w$, regardless of the value of $Z_w$.
 This rather odd situation calls into question the meaning of
\textit{active} and what we might realistically expect can be inferred from data.
Indeed, the issue is related to identifiability of the activity vector $Z$, since it
shows that distinct $Z$ vectors may produce the same part-level activity vector
$A= (A_p )$.
(In the case above, switching $Z_w$ from 0 to 1 does not change $A$.) The
mapping $Z \longrightarrow A$ given by~(\ref{eqf1}) is not necessarily invertible,
depending on the system as defined in $I$.   Lack of identifiability would not
necessarily create difficulty in a Bayesian analysis, however, in the present case
we are specifically interested in inferring the activity states of the gene sets
and prioritizing these sets, and so it stands to reason that we ought to confer a real,
if still only model-based, meaning on the activities.

When \textit{activity} is defined more fully, there is a simple solution to the problem.
The \textit{activation hypothesis} asserts that a set of parts is active if and only
if all parts in the set are active.  It was shown previously (Newton et al.) as follows:

\begin{thm}
\label{thmah}
Under the activation hypothesis ($\mbox{AH}$), the mapping $Z \longrightarrow A$ defined by
\[
A_p = \max_{w\dvtx  p \in w} Z_w
\]
is invertible, with inverse $A \longrightarrow Z$
\[
Z_w = \min_{p\dvtx p \in w} A_p.
\]
\end{thm}

The inverse mapping is simply that a whole is inactive if and only if
any of its parts is inactive.
So the odd case at the beginning of the section cannot occur under AH;
if all parts are active, then $Z_w=1$ must hold.  Further, with
parameters~$\alpha$ and~$\gamma$ fixed, the $Z$ vector is identifiable under AH,
since different $Z$ vectors necessarily give different probability distributions
to data $X$.

The first contribution of the present work is to show that the activation
hypothesis is equivalent to a set of linear inequality constraints on the activity
variables. The finding is useful for posterior inference computations.
We prove in Section~\ref{sec7} the following:

\begin{thm}\label{thm22}
$\mbox{AH}$ holds if and only if all of the following hold:
\begin{longlist}[3.]
\item[1.] $Z_w \leq A_p$ for all $p, w$ with $p \in w$;
\item[2.] $A_p \leq \sum_{w\dvtx p \in w} Z_w$ for all $p$;
\item[3.] $\sum_{p\dvtx p \in w}  ( Z_w - 2A_p + 2  ) \geq 1$ for all $w$.
\end{longlist}
\end{thm}

Evidently, the i.i.d. Bernoulli prior (\ref{eqprior1}) does not respect AH in the
sense that vectors $Z$ which violate AH have positive prior probability.  In
simple systems such violation may be innocuous. We
provide evidence that in the complex systems such as GO,
this violation creates a substantial
loss of statistical efficiency. We note first that alternative prior specifications
are available that respect AH.  A simple one is to condition prior (\ref{eqprior1})
on the AH event, namely,
\begin{equation}
\label{eqprior2}
P ( Z =z ) = \biggl( \frac{1}{c} \biggr) \pi^{\sum_{w} z_w}
(1-\pi)^{ \sum_w (1-z_w)} \qquad  \mbox{if $z$ satisfies AH},
\end{equation}
otherwise $P(Z=z)=0$,  where $c$ is the probability,
in prior (\ref{eqprior1}), that $Z$ satisfies AH, and $z$ is a vector of binaries
representing a possible realization of $Z$.  In other words, with subscript ``1''
for the i.i.d. prior (\ref{eqprior1}) and ``2'' for prior (\ref{eqprior2}), we
have
$ P_2 ( Z=z  ) = P_1 ( Z=z | \mbox{AH})$.
Upon conditioning, the $(Z_w )$ are not necessarily either
 mutually independent or identically distributed.

\section{Statistical properties}

The role of the prior distribution in Bayesian analysis has surely been the subject
of considerable debate.  On the one hand, it helps by regularizing inference, especially
in high dimensions.  On the other hand, data need to work against it to produce inferences
that trade off empirical characteristics with prior assumptions.  A fact of relevance
to the present problem is that gene-list data must work against either prior
[(\ref{eqprior1}) or (\ref{eqprior2})] to deliver an inferred list of activated gene sets.
 For
two Bayesian analysts, one using prior (\ref{eqprior1}) and the other using
prior (\ref{eqprior2}), the true state is ascribed different prior mass.  The ratio
of these masses, $\rho$, represents the extra effort needed to be done by the data to overcome
prior (\ref{eqprior1}) compared to prior (\ref{eqprior2}):
\begin{equation}
\label{eqratio}
\rho = \frac{ P_2 ( Z= z_{\mathrm{true}}  ) }{ P_1 ( Z= z_{\mathrm{true}})} = \frac{ P_1 ( Z=z_{\mathrm{true}} |
{\mbox{AH}})}{ P_1 (
  Z= z_{\mathrm{true}})} = \frac{1}{P_1 ( {\mbox{AH}})}
\geq 1.
\end{equation}
Here we have used the particular structure of prior (\ref{eqprior2}) and also
the assumption that $z_{\mathrm{true}}$ satifies AH.
If $z_{\mathrm{true}}$ did not satisfy AH, the target of inference would be beyond the realm
of any gene-level data set to estimate, owing to lack of identifiability. Indeed, it is
difficult to see what meaning could be ascribed to $z_{\mathrm{true}}$ in that case.
The observation to be gained from (\ref{eqratio}) is that the probability of AH under
the i.i.d. prior affects the efficiency of inference.  In systems where
that probability is very small, there is reason to believe that improved inferences
are possible.  As to the precise effect
of ignoring AH,\vadjust{\goodbreak} that depends on the particular system $I$, the true activation state,
and the system parameters $\alpha$ and $\gamma$.  What our initial investigation finds
is that a  truly activated whole $w$  may tend to have
 $P_1( Z_w=1 | X)$  smaller than $P_2( Z_w=1 | X)$, and if so the $P_1$ inference is
too\break conservative.


Whether or not AH holds for a given state $Z$ may be assessed by calculating the part-level
activities $A$ and then checking Theorem~\ref{thm22}. Alternatively, we consider whole-level
\textit{violation} variables $(V_w)$.  These Bernoulli trials are defined as follows:
\begin{equation}
\label{eqviolation}
V_w = \cases{1, &\quad  \mbox{if $Z_w=0$ and if for all $p \in w$ there exists $w'$}
\cr
& \quad \mbox{with $p\in w'$ and $Z_{w'}=1$,}
\cr
0, &\quad \mbox{otherwise}.}
\end{equation}
The probability, under $P_1$, that $Z$ satisfies AH is equivalent to the probability
of no violations, that is,
\begin{equation}
\label{eqviolate}
P_1(\mbox{AH}) = P_1 ( V_w = 0,
\forall w ),
\end{equation}
and so AH probability might be approachable by considering the violation variables.
Except in stylized examples, we do not expect these variables to be mutually independent;
indeed, they may have a complicated dependence induced by overlaps of the wholes and,
hence, direct calculation of $P_1(\mbox{AH})$ is intractable.
However,
the expectations of $V_w$ are readily computable for a given system, either by Monte Carlo or
by a more sophisticated algorithm
 [\citet{wa14}]. Considering the Chen--Stein result
for Poisson approximations,
we conjecture that $-\log P_1(\mbox{AH})$ is approximately equal to $E_1 ( \sum_w V_w  )$,
though we have not been able to guarantee an error on this approximation
[cf. \citet{ar90}].

\begin{figure}

\includegraphics{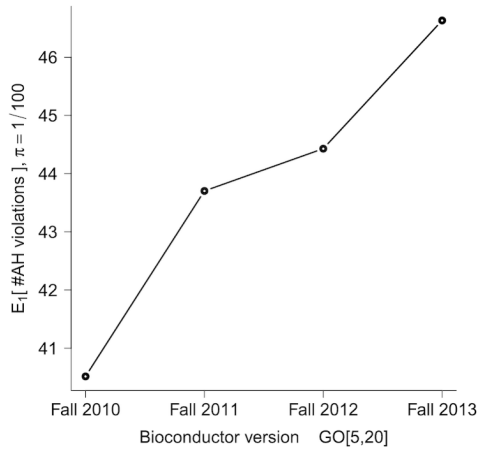}

\caption{Expected number of sets that violate the activation hypothesis (AH) for four recent versions of Gene
 Ontology (GO), considering sets holding between 5 and 20 genes, taken on the i.i.d. Bernoulli prior. Calculations
 are done at $\pi = 1/100$.  Respectively, these systems contain 3591, 4096,
 4449 and 4772 gene sets, and correspond to versions of org.Hs.eg.db.}\label{figgo}
\end{figure}

Figure~\ref{figgo} charts the expected value  $E_1 ( \sum_w V_w  )$
 over four recent versions
of Gene Ontology, for $\pi = 1/100$.  For concreteness it focuses on GO terms holding
between 5 and 20 genes (for which an exact calculation of the expectation is feasible),
though the key finding is not sensitive to that restriction, as evidenced by Monte
Carlo computations (not shown).  As one might expect by the increasing density and
complexity of GO,
the expected number of AH violations increases. This may very well reflect the
fact that $P_1(\mbox{AH})$ is decreasing over time,   which indicates to us that ignoring AH
is becoming an ever greater problem for gene-set analysis.


In terms of modeling assumptions, there is no additional cost to accounting for AH
in the Bayesian analysis; the cost is purely computational, since inference must now
deal with the constraints imposed by AH on the space of latent activities.  The next
sections describe two computational advances that address the problem.\vadjust{\goodbreak}

\section{Decoding functional signals via constrained optimization}

\subsection{MAP via ILP}

Decoding a discrete signal is frequently accomplished by algorithms that compute
the parameter state having the highest posterior mass: the maximum
 a posteriori (MAP) estimate.
Although limited as a posterior summary, the MAP estimate may contain
 useful multivariate information  [e.g., \citet{cl08}].
Our representation of model-based gene-set analysis reveals that
 under model (\ref{eqmodel}) and prior (\ref{eqprior2}),
the log posterior is linear in the joint collection of whole and part activity variables.
This log posterior is
\begin{eqnarray}
l(Z,A) &=& \sum_w \bigl\{
Z_w \log( \pi ) + (1-Z_w) \log( 1- \pi) \bigr\}
\nonumber\\
\label{eqobjective}
&&{}+\sum_p \bigl\{
A_p \bigl[ x_p \log(\gamma) + (1-x_p)
\log(1-\gamma) \bigr]
\\
\nonumber
&&\qquad\hspace*{9pt}{}+ (1-A_p) \bigl[ x_p \log(
\alpha) + (1-x_p) \log (1-\alpha) \bigr] \bigr\},
\end{eqnarray}
where $x_p$ is the realized value of the gene-list indicator $X_p$, and $\alpha$, $\gamma$,
and $\pi$ are system parameters, which are considered fixed in the present
calculation.  Considering Theorem~\ref{thm22},
finding the MAP estimate $(\hat{Z}, \hat{A})$ amounts to maximizing
a linear function in discrete variables
subject to linear inequality constraints.  As such, it fits
naturally into the domain of \textit{integer linear programming} (ILP), an
active subfield of optimization. Our computations
take advantage of ILP software available in the GNU
Linear Programming Kit through its interface
with \texttt{R}.\footnote{See \surl{www.gnu.org/software/glpk} and
\surl{cran.R-project.org/package=Rglpk}.}
We employed
a series of basic code checks to assure our implementation worked in: (1)
simple examples where the MAP estimate is computable by other means; and (2)
limiting situations where $X_p$ was Binomial having a high sample size, and
thus where the MAP estimate must converge to the true activity state.

The reconstruction $\hat{Z}$ obtained through this optimization holds an
estimate of the activated and inactivated gene sets.  We refer to the overall method
as the \textit{multi-functional analyzer} (MFA), and specifically MFA-ILP to refer
to the posterior mode computed by ILP.  We note that by invertibility of the mapping
 $Z \longrightarrow A$ under AH,  the log-posterior $l$ could be expressed either as
 a function of $Z$ only or as a function of $A$ only, however, in neither reduced
case would $l$ be linear in the input variables.  Moreover, in neither reduced case
could the constraints be expressed as linear inequality constraints.
 By expanding the domain
we formulate the constrained optimization as an integer linear program.

\subsection{Numerical experiments}

In each experiment reported below we represented a system with a
parts-by-wholes incidence matrix $I$; we fixed the false-positive rate
$\alpha = 1/10$ and the true positive rate $\gamma = 9/10$.  We simulated 100 gene-lists $X$ from
model (\ref{eqmodel}), each time using  a simulated  activity vector $Z$.
For methods comparison, we applied the following: (1) the commonly used Fisher exact test
for enrichment of each gene set in the data $X$ [\citet{k12}], (2) MGSA (version 1.7.0), and (3) MFA-ILP.
We allowed both model-based methods to know the system parameter settings.
To evaluate performance,
we calculated specificity, sensitivity, and precision of the estimated
activity vector~$\hat{Z}$ for the true activity vector~$Z$ by averaging over
the 100 replicates.

\subsubsection*{Experiment  1: Low overlap}
Initially, $I$  had
size $300$ genes (parts)  by $100$ gene sets (wholes).
 We randomly picked 5 and 10 parts for each whole in columns 1--50 and 51--100,
 respectively. Then we removed parts not contained by any whole,
 leaving a $296 \times 100$ incidence matrix.
 We sampled $Z$ from prior (\ref{eqprior1}) and then projected it onto
 AH by constructing $A_p=\max_{w\dvtx p \in w} Z_w$ and then updating
 $Z_w= \min_{p\dvtx p \in w} A_p$.
   All methods exhibit similar operating characteristics in this case
 (Table~\ref{tablow}).

\begin{table}
\caption{Simulation comparison of three gene-set methods, a case of low overlap
 among gene sets: Tabulated are mean values from 100 simulated data sets. On average
 7.3 truly activated sets occur}\label{tablow}
\begin{tabular*}{\tablewidth}{@{\extracolsep{\fill}}lcccc@{}}
\hline
\textbf{Method}&   \textbf{Predicted \# active} &\textbf{Sensitivity}&\textbf{Specificity}&\textbf{Precision}\\
\hline
MFA-ILP&7.4              &0.963 &0.997 &0.958\\
Fisher (cut-off${}={}$0.05)&5.9&0.790 &0.998 &0.966\\
Fisher (cut-off${}={}$0.1)&6.8 &0.873 &0.996 &0.948\\
MGSA (cut-off${}={}$0.5)&7.2   &0.954 &0.998 &0.968\\ \hline
\end{tabular*}
\end{table}

\subsubsection*{Experiment 2: Higher overlap and parent-child structure}
 Initially, $I$ had size $300$ parts by $105$ wholes.
 From column 1 to column 20, each column has 20 parts, of which 15 parts
 are in common with each other and 5 parts are randomly selected from the
 other parts; column 21 has 10 parts which are randomly picked from the
 15 common parts shared by columns 1--20. Thus, columns 1--20 have a lot of
 overlaps and column 21 is a child of columns 1--20.
 Similarly, we built columns 22--42, 43--63, 64--84 and 85--105.
 The common 15 parts in each column combination are all different.
 Then parts not contained by any whole were removed, which resulted in a
 $265\times 105$ incidence matrix.
 We activated wholes by sampling one whole from columns 1--20, 22--41,
 43--62, 64--83 and 85--104 as activated, respectively, and projected onto AH as
 above.

\begin{table}[b]
\caption{Simulation comparison of three gene-set methods, a case of higher overlap
 among gene sets:  Tabulated are mean values from 100 simulated data sets.
 On average there are 10.1 truly activated sets in this case}\label{tabhigh}
 \begin{tabular*}{\tablewidth}{@{\extracolsep{\fill}}lcccc@{}}\hline
\textbf{Method}&  \textbf{Predicted \# active} &\textbf{Sensitivity}&\textbf{Specificity}&\textbf{Precision}\\
\hline
MFA-ILP&\phantom{0}10.2              &0.975 &0.997 &0.993\\
Fisher (cut-off${}={}$0.05)&104.2&0.996 &0.008 &0.096\\
Fisher (cut-off${}={}$0.1)&104.8 &0.996 &0.002 &0.096\\
MGSA (cut-off${}={}$0.5)&\phantom{00}5.5    &0.490 &0.995 &0.920 \\
\hline
\end{tabular*}
\end{table}

  Table~\ref{tabhigh} exhibits properties of three methods
in relatively complicated system just defined.
   The univariate Fisher test tends to select the
wholes with a high correlation (overlap) with the truly activated wholes,
 which results in high sensitivity but low specificity (or precision).
The extra activation calls correspond to spurious associations that the multivariate,
model-based approaches are able to recognize.
The MGSA method often fails to discover truly activated wholes, which
corresponds to a reduced sensitivity.  The small \textit{child} wholes tend to
be missed by MGSA in this case.  The proposed MFA-ILP method is right on target.

\subsection{ILP for large systems}

Large systems strain unaided ILP computation, but the special structure of the gene-set problem
allows for several refinements.

\subsection*{Shrinking $I$}  Up to a constant, the objective function in~(\ref{eqobjective})
may be expressed
\[
l(Z,A) = c_1 \sum_{w}
Z_w + c_2 \sum_{p \in P^{-}}
A_p + c_3 \sum_{p \in P^{+}}
A_p,
\]
where
\begin{eqnarray*}
c_1 &=& \log \pi - \log(1-\pi),
\\
c_2 &=& \log (1-\gamma) - \log( 1-\alpha),
\\
c_3 &=& \log \gamma - \log \alpha
\end{eqnarray*}
and where $P^-$ and $P^+$ denote the observed inactivated and activated parts,
respectively.  That is, $p \in P^-$ if $x_p=0$ and $p \in P^+$ if $x_p=1$.
By assumption~(\ref{eqmodel}), $\alpha < \gamma$ and so $c_2 <0$ and $c_3 > 0$.
If we further insist that $\pi < 1/2$, then $c_1 < 0$ also.  In some cases we
can know which $\hat{Z}_w$ and $\hat{A}_p$
  must equal 0 in the optimal solution, and if so we can remove
these variables from the system prior to implementing ILP.  For each whole $w$ denote
$P^{+}_{w} = w \cap P^+$  and similarly $P^{-}_{w} = w \cap P^-$, and define
$W^* =  \{ w\dvtx c_1 + c_3 \sum_{p \in P^{+}_{w}} 1  < 0  \}$.
  Clearly, those
wholes containing no reported parts are in $W^*$, but there may be others.  We prove
in Section~\ref{sec7} that if $W^*$ is not empty, then we may be able to shrink the system prior to
solving the constrained optimization problem via ILP.

\begin{thm}\label{thm41}
Suppose $\pi < 1/2$ and let $w_0$ denote an element of $W^*$.  If there exists $p_0 \in w_0$ such that
$\{ w\dvtx  p_0 \in w \} \subset W^*$, then $\hat Z_{w_0}=\hat A_{p_0}=0$.
\end{thm}

Letting $W_0$ and $P_0$ denote wholes and parts for which the optimal solution is known
(in advance of computation), we may remove these from the incidence matrix $I$,
effectively shrinking it.  The amount of shrinkage may be dramatic, but it depends on
the observed data $x$, the system $I$, and system parameters $\alpha$, $\gamma$ and
$\pi$.  When $\alpha$ is small and $\gamma$ is large, the effects may be minimal.

\subsubsection*{A sequential approach} In the unlikely event that the system matrix $I$ is
 separable into blocks of wholes that do\vadjust{\goodbreak} not overlap between blocks,
then ILP may be applied separately
 to these distinct blocks in order to identify the MAP activities.
 We do not expect this separability in GO or related systems,
but we can take advantage of size variation of the wholes and work sequentially
from small ones to larger ones.
As an example, let $S_{10}$ denote the sets containing no more than $\mathit{n.up}=10$ genes.
In order to obtain the optimal solution for the full problem, we
start from the sub-matrix $I.10$ obtained by extracting these sets from $I$.
Suppose $Z^*_{10}$ is the MAP solution based on the data for $I.10$, and use  notation
$S^*_{10}$ to denote the active sets in $S_{10}$ as inferred by $Z^*_{10}$.
We aim to find the optimal solution $Z^*_{11}$ for $I.11$ using what has\vspace*{1pt} already been
computed in the smaller system.  Denote the newly added sets in $I.11$ by
 $S_{10}^{11}$ (i.e., the sets containing exactly 11 genes).
We just need to consider the sets with the possibility being active in the optimal
solution on~$I.11$. First of all, $S^*_{10}$ and $S_{10}^{11}$ should be included,
in the case we have no any other prior knowledge about $Z^*_{11}$. Second,
by the 3rd AH inequality (Theorem~\ref{thm22}), any set in $S_{10}\setminus S^*_{10}$
which is a subset of some set in $S_{10}^{11}$, denoted by $D$, should also be included.
But these sets already considered are not enough. Actually, for each set $w_1$
in $S_{10}^{11}$, we need to check whether there exists some set
$w_2$ in $S_{10}\setminus (S^*_{10}\cup D)$ satisfying
\begin{equation}
\label{eqsequential} c_1 + c_2 \sum
_{p\in P^{-}\cap P_{w_1}^{w_2}}A_p+ c_3\sum
_{p\in P^+\cap P_{w_1}^{w_2}}A_p>0,
\end{equation}
where $P_{w_1}^{w_2}$ denote the set of genes contained by $w_2$ and not by $w_1$.
We do this since each set in $S^{11}_{10}$ may be active in the optimal
solution $Z^*_{11}$, and we need to check whether some sets in $S_{10}$ should be
activated toward maximizing the objective function. We denote the sets
in $S_{10}\setminus (S^*_{10}\cup D)$ satisfying the condition~(\ref{eqsequential})
 by~$E$.  Finally, by the 3rd AH inequality (Theorem~\ref{thm22}), any set in
$S_{10}\setminus (S^*_{10}\cup D\cup E)$ which is a subset of some set in $E$,
denoted by $F$, should also be included. Thus, we need to run the ILP on the
incidence matrix only for $S^*_{10}\cup S^{11}_{10}\cup D  \cup E\cup F$,
instead of $I.11$.
Hence, we obtain a sequential approach to solve the full ILP problem from a sequence
of smaller problems. Examples show this is feasible in GO for subsystems holding sets of
up to 50 genes, without excessive computational burden.

\section{Posterior sampling}
\subsection{Penalized MCMC}

To obtain a sample from the posterior distribution defined by prior~(\ref{eqprior2}) and model~(\ref{eqmodel})
in which the whole activity variables $Z = (Z_w)$ have positive probability only when $Z$ satisfies
AH, we design a Markov chain to run within the unconstrained space according to a penalized posterior:
\begin{equation}
\label{eqpenalizedMCMC}
\tilde{l}(Z) = l(Z,A) - \lambda \sum
_w V_w,
\end{equation}
where $l(Z,A)$ is defined in (\ref{eqobjective}), $V_w$ is the violation
 indicator~(\ref{eqviolation}), and $\lambda \ge 0$ is a tuning parameter.
The desired sample is obtained by discarding any sampled states that do not satisfy AH.
Note that there are no violations ($\sum_w V_w = 0$) for $Z$ that satisfy AH,
so that $\tilde{l}(Z) = l(Z,A)$ in this case
and the conditional log posterior distribution under $\tilde{l}(Z)$ restricted to AH is identical to the target log posterior distribution.
Increasing the tuning parameter $\lambda$ increases the probability of AH in the larger state space,
which is essential for efficient sampling when this probability is small.
We find that penalizing the log posterior within
 the unconstrained space leads to a conditional sampler that mixes well in the constrained space,
where our previous attempts to constrain move types  were less successful.

It is helpful to visualize the Markov chain as operating by changing colors
on the node-colored bipartite graph $\mathcal{G}$
having whole nodes and part nodes,
with an edge between a whole node $w$ and part node $p$ if and only if $p \in w$,
and where the coloring of the whole nodes $\{w\}$ and part nodes $\{p\}$ match the activities $Z$ and $A$, respectively.
It is useful in assessing the state of the Markov chain to associate with each node a count $n(\cdot)$ of its active connected neighbors in $\mathcal G$.
$A_p=1$ if and only if $n(p)>0$ and $V_w = 1$ if and only if $Z_w = 0$ and $n(w) = \deg(w)$,
the number of part nodes $p \in w$.

The Markov chain proceeds by selecting at random a whole node $w$
and proposing a color swap (a change in the status of the activity variable, $Z_w^* = 1 - Z_w$) for this node.\footnote{Efficiency gains may be possible
by using a nonuniform sampler, for instance, depending on the set size or the number of reported parts in the whole, though we use a uniform
proposal in the present work.}
This proposed change can, but need not, affect the activities of parts contained in this whole.
When $Z_w^* = 1$,
the active neighbor counts $n(p)$ increase by 1 for each $p \in w$.
If $A_p$ changes from 0 to 1,
then each node $w^\prime$ that contains $p$ (including $w$) gains an additional active neighbor and $n(w^\prime)$ increases by 1.
This increase could cause a violation if $p$ were the only remaining inactive neighbor of an inactive $w^\prime$, causing $V_{w^\prime}$ to change form 0 to 1.
If node $w$ were in violation before this proposal,
activating it would eliminate the violation.
Similarly,
when $Z_w^* = 0$,
the active neighbor counts $n(p)$ decrease by 1 for each $p \in w$.
If this decrease is from 1 to 0, then the activity $A_p$ changes from 1 to 0 as well
and all of the whole nodes $w^\prime$ connected to $p$ would lose an active neighbor, $n(w^\prime)$ decreasing by 1.
If the whole node $w^\prime$ had been in violation,
this change would eliminate the violation with $V_{w^\prime}$ changing from 1 to 0.

\begin{figure}[b]

\includegraphics{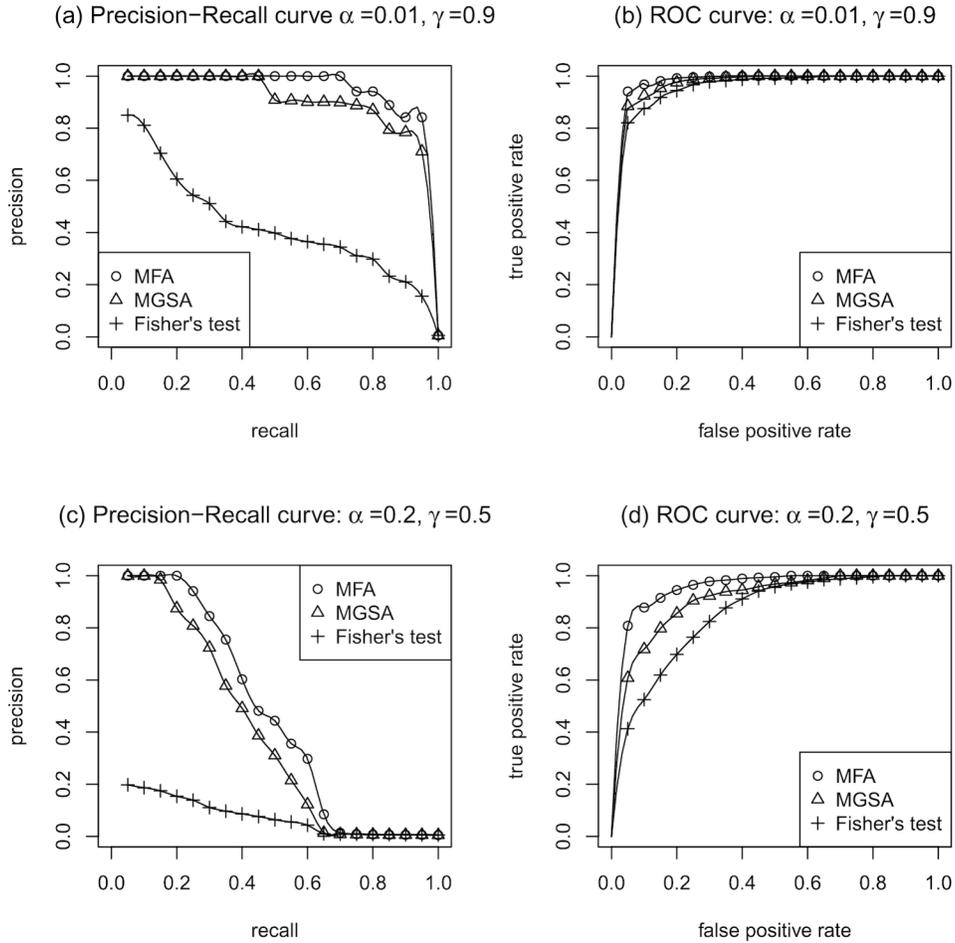}

\caption{Operating characteristics of MFA-MCMC, MGSA, and Fisher's test based on simulating the role
 model in the \textup{D. melanogaster} genome.}\label{figsim}
\end{figure}

Careful accounting of the changes to a few key counts allows for quick calculation of the change in $\tilde{l}(Z^*)$
and subsequent\vspace*{1pt} acceptance or rejection of the proposal by Metropolis--Hastings.
The log posterior $\tilde{l}(Z^*)$ is a function of $\alpha$, $\gamma$, $\pi$,
the penalty $\lambda$
and the counts of the numbers of active and inactive whole nodes [$\sum_w Z_w$ and $\sum_w (1-Z_w)$, resp.],
the number of whole nodes in violation ($\sum V_w$),
the numbers of active part nodes with realized values 1 and 0 [$\sum_p A_p x_p$ and $\sum_p A_p (1-x_p)$, resp.],
and the numbers of inactive part nodes with realized values~1 and 0 [$\sum_p (1-A_p) x_p$ and $\sum_p (1-A_p) (1-x_p)$, resp.].

\subsection{Numerical experiment}

To assess the performance of MFA-MCMC, we simulated gene-list data
according to the role model in the \textit{D. melanogaster} genome, following the scheme presented in \citet{b10}.
 Briefly,  we used 3275 GO terms annotating between 5 and 50 fly genes, according to version 2.14.0 of
Bioconductor package org.Dm.eg.db. In each simulation run, a number of GO terms were activated and then
a gene list was constructed from independent Bernoulli trials depending on the activation states and settings
of false-positive and false-negative error rates.  Figure~\ref{figsim} shows receiver-operating (ROC) curves
and precision-recall curves for two parameter settings, based on 100 simulated gene lists in each setting.
   Selection to the reported set list is based on
thresholding the marginal posterior probability (MGSA, MFA-MCMC) or the \mbox{$p$-value} (Fisher). Evidently, MGSA and MFA-MCMC
are accurate and show similar behavior when error rates are low,
 though MFA-MCMC shows improved precision and sensitivity in more difficult settings.

In subsequent calculations we deploy both MAP estimation (MFA-ILP) and MCMC sampling on each data set in order
to infer wholes that are probably activated.  For MCMC, we use $10^7$ sweeps, burn-in of $10^6$, and $\lambda=5$,
which causes about one third of the states
 to satisfy $\mbox{AH}$. MFA-ILP gives a summary functional decoding
of the gene list.  Posterior probabilities from the MCMC computation provide a measure of confidence in the
inferred sets and also highlight notable non-MAP sets.  Fisher's test is the default univariate method for gene-list
 data;  we include it for comparison, even though the hypotheses it tests are different from the
 activation states assessed by MFA and MGSA.

\section{Examples}

\subsection{Genes implicated in type 2 diabetes (T2D)}
From a large-scale genome-wide  association study (GWAS) involving more than 34,000 cases and 114,000 control subjects,
77 human genes have been implicated as affecting T2D disease susceptibility [\citet{morris}, Supplementary
Table~15, primary list].  To assess the functional content of this gene list, we applied MFA, MGSA, and simple
enrichment via Fisher's exact test, all in the context of 6037 gene ontology terms, each annotating between 5 and 50
genes.\footnote{These 6037 terms annotate a total of 10,626 genes; among the 77 T2D-associated genes, 58 are  in this
\textit{moderately annotated} class.}
Here and in other examples we took advantage of available information on likely false positive ($\alpha$) and
false negative $(1-\gamma)$ error rates at the gene level.  Using the fitted mixture model from Morris et al., we estimated
$\alpha = 0.00019 $ and $\gamma = 0.02279 $ for this large-scale GWAS [details in \citet{wa14}].

\begin{figure}

\includegraphics{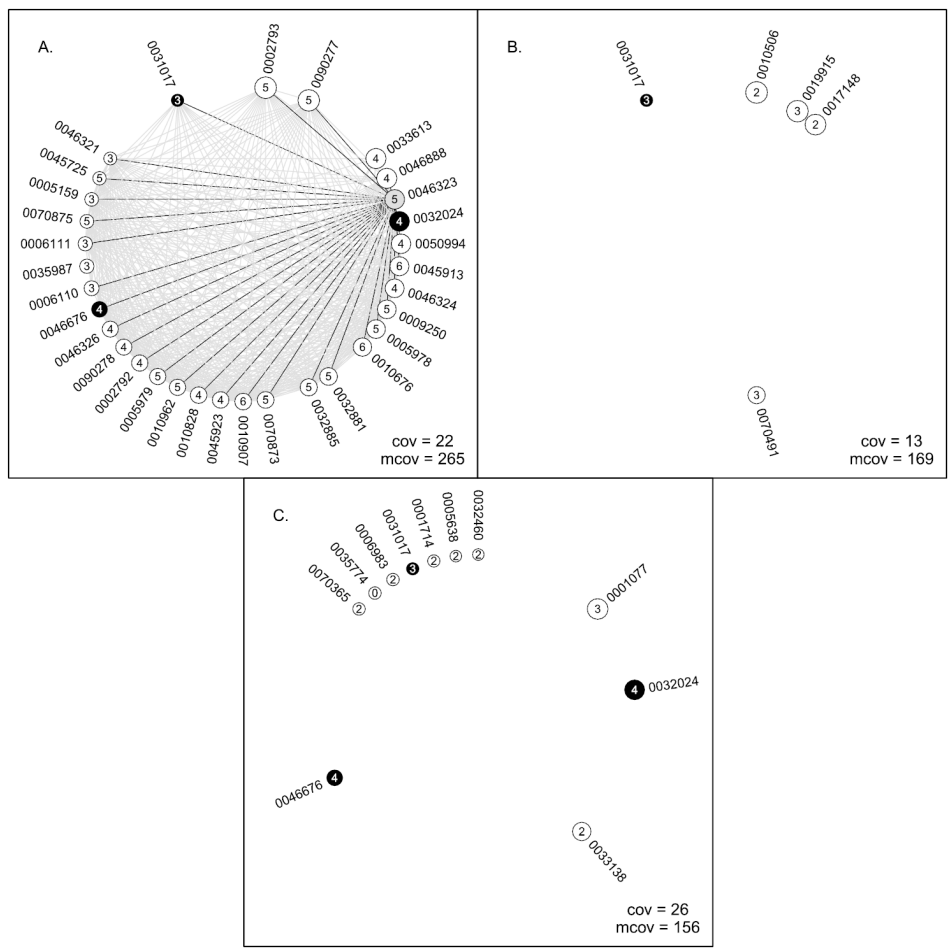}

\caption{GO terms inferred by three methods (\textup{A}, Fisher's test; \textup{B}, MGSA; \textup{C}, MFA-ILP)
 to be activated to explain
 the type 2 diabetes associated genes (58 of the 77 are annotated to the 6037 GO terms used).
  Panels are linked by the location of sets; three sets (black) were identified by multiple methods. GO ID is on
 the outer rim, and sets are ordered by size as in Tables~\protect\ref{tabT2Dmfa} and \textup{S1--S3},
 from largest (noon) and decreasing clockwise;
  numbers inside circles indicate the number of T2D genes annotated to that set;
 a line connects two sets if their intersection contains at least one T2D-associated gene.
  The total number of T2D genes explained by
the identified sets (coverage), lower right, as a subset of the 58 starting genes.
 Similarly, the mis-coverage (mcov) counts the number of genes within the inferred active
 sets that are not on the observed gene list.
 GO:0046323 (glucose import) and its relations are highlighted in panel \textup{A} (grey circle, black lines).\vspace*{-6pt}}\label{figt2d}
\end{figure}

\begin{table}
\caption{MFA results in type 2 diabetes (T2D) example:  11 GO terms are inferred to be active
 using the ILP algorithm to compute the MAP estimate (rows).  Basic statistics on these terms are provided   (\#~T2D-associated
genes/set size).  The next two columns give the MCMC-computed  marginal posterior activation
 probabilities for these terms, both using MGSA and MFA, the constrained alternative.  The final column holds the Benjamini--Hochberg adjusted Fisher-test
$p$-value.  All calculations start with 6037 GO terms (those annotating between 5 and 50 human genes) that together annotate 10{,}626 human genes.
   Of the 77 total T2D genes, 58 have at least one annotation to these 6037 GO terms.
 The inferred gene sets cover 26 of these 58 T2D genes}\label{tabT2Dmfa}
\begin{tabular*}{\tablewidth}{@{\extracolsep{\fill}}lcccc@{}}\hline
\textbf{Gene set (GO term)}&\textbf{Statistics}&\textbf{P.MFA}&\textbf{P.MGSA}&\textbf{Fisher} \\
\hline
RNA polymerase II core promoter$\ldots$ & $3/45$ &0.517&0.028&0.161 \\
Positive regulation of insulin secretion& $4/41$ &0.964&0.372&0.016 \\
Positive  regulation of peptidyl-serine$\ldots$ & $2/35$ & 0.537&0.096&0.756 \\
Negative  regulation of insulin secretion& $4/23$&0.996&0.201&0.003 \\
ER overload response&\phantom{0}$2/9$ & 0.398 & 0.159 & 0.102 \\
Positive regulation of insulin secretion$\ldots$ & \phantom{0}$0/9$&0.964&0.002&1.000 \\
Hepatocyte differentiation&\phantom{0}$2/9$ &0.316&0.016&0.102 \\
Endodermal cell fate specification& \phantom{0}$2/8$& 0.596 & 0.036 & 0.091 \\
Exocrine pancreas development&\phantom{0}$3/8$ & 0.946 & 0.600 & 0.003 \\
Negative regulation of protein$\ldots$ & \phantom{0}$2/5$ & 0.420 & 0.101 & 0.051 \\
Lamin filament& \phantom{0}$2/5$ & 0.790 & 0.400 & 0.051 \\
\hline
\end{tabular*}
\end{table}

Figure~\ref{figt2d} summarizes the application of MFA, MGSA, and Fisher's test to this example.
 Table~\ref{tabT2Dmfa} reports those gene sets
 inferred by MFA-ILP to be activated in T2D. Tables S1--S4 provide further
 information  for comparison of MFA with MGSA and Fisher's test
 [\citet{wa14}].
The example illustrates features we see repeatedly with these methods.   Sets identified
by Fisher's test tend to overlap substantially, reflecting the univariate nature of the approach;
both MGSA and MFA-ILP alleviate this redundancy problem, but MGSA finds fewer sets than MFA-ILP.
 As expected, each of the 11 sets inferred by the ILP algorithm (i.e., the MAP
estimate of the activated sets) has high marginal
 posterior probability of activation (P.MFA).  Furthermore, MFA-ILP is able to explain more
of the gene level findings than the other methods, as indicated by the number of genes
 that are both in the reported gene list (T2D) and are in at least one of the gene
 sets inferred to be activated (\textit{coverage}). It does this without increasing the
 mis-coverage, which is the number of non-T2D genes within the inferred active sets.
  Figure~\ref{figt2d} summarizes the sets found by these
 methods and reports these coverages.

 An interesting set in this case is the GO term  \textit{glucose import} (GO:0046323),
 for which the proportion $5/41$
 of observed T2D genes is very high (small Fisher \mbox{$p$-value}), but there is very small posterior
activation probability according to MFA. That is because
 the 5 genes are explained more easily as parts of three other terms in the MAP estimate
 that have yet other genes supporting their activation.

 A second curious case is GO:0035774, a small term (9 genes) to do with regulation of insulin
 secretion.  None of these genes was reported to be involved in T2D, however, the set is fully
 contained in a parent set  which is inferred to be activated by MFA-ILP.  As the calculation
 respects AH, all subsets of activated sets are activated.  This may be a set-level false-positive call,
as none of the contained genes was reported to be T2D associated.  MFA favors the explanation that each
 of the 9 noncalls was a gene-level false negative, finding the weight of evidence supporting that interpretation.
 When we recall  that the
 gene-level false-negative rate is almost 98\% (following the mixture calculation from
 Morris et al.), this assessment seems plausible. We note that for the sake of further
 simplification of output, it is reasonable to suppress any such subsets from primary
 tabulations [see trimming algorithm, \citet{wa14}].

\subsection{RNA interference  and influenza-virus replication}

\begin{figure}

\includegraphics{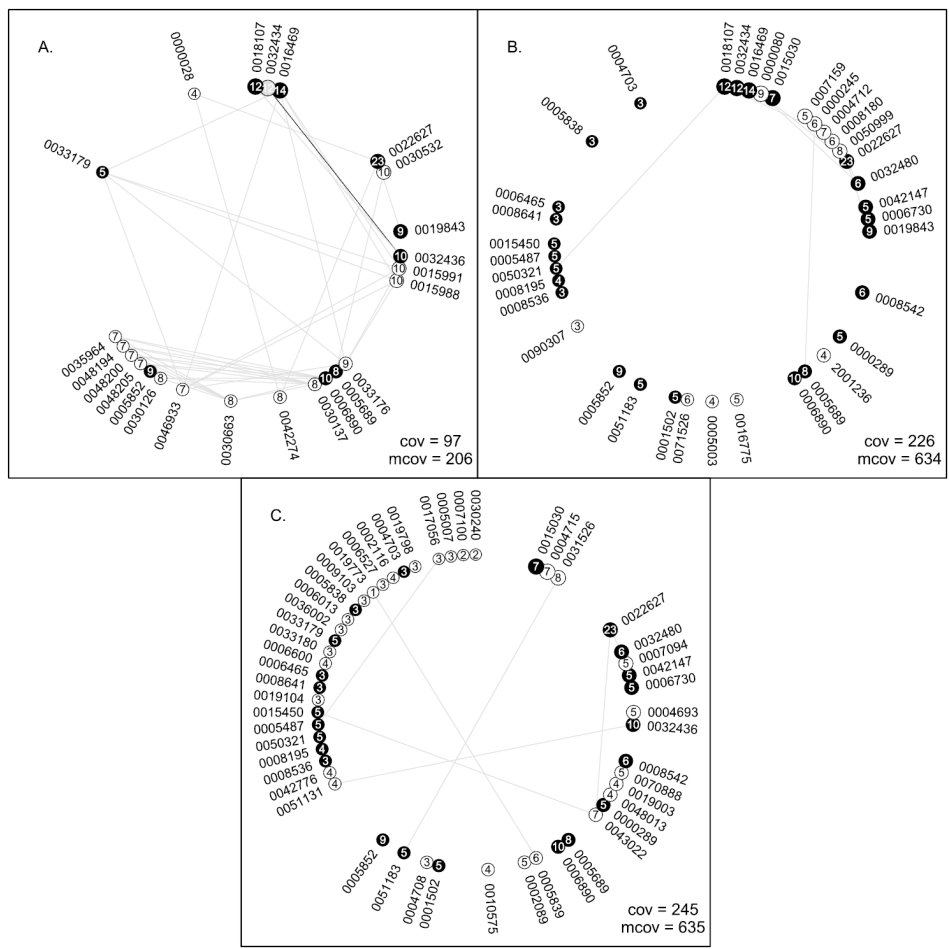}

\caption{Sets (GO terms) identified by three methods (\textup{A}, Fisher's test;
 \textup{B}, MGSA; \textup{C}, MFA) as activated to explain
 the influenza-involved human genes. The layout is as in Figure~\protect\ref{figt2d}.}\label{figflu}
\end{figure}

In a meta-analysis of four genome-wide studies of influenza virus, \citet{metaflu}
reported that 984 human genes had been detected by RNA interference as possibly being associated
with viral replication. As in the T2D example, we compared MFA with MGSA and Fisher's test
 on this gene list using 6037 GO terms annotating between 5 and 50 human genes.  Among the 984
 influenza-involved genes, 683 are annotated to at least one of these terms. To apply the
model-based methods, we took advantage of external information on the false positive rate $\alpha$
and the true positive rate $\gamma$ [see \citet{wa14}].
  Figure~\ref{figflu} illustrates the sets
found by MFA-ILP, MGSA, and Fisher's test, and Tables S5--S8 [\citet{wa14}] contain further details
 of the comparative analysis.  Again, we find that MFA-ILP dominates the other methods in terms
 of gene coverage, with 245 genes explained as compared to 226 (MGSA) and 90 (Fisher) and
 with mis-coverages 635 (MFA-ILP), 634 (MGSA), and 206 (Fisher).
 Furthermore, MFA-ILP detects more sets than MGSA (50 in the trimmed list compared to
 30 by MGSA).

  To better understand differences between the methods, consider one gene set GO:0032434, \textit{regulation of proteasomal
  ubiquitin-dependent protein\break catabolic process}, which annotates 48
 human genes, 12 of which were in the 984-list of influenza involved genes (highlighted in
 Figure~\ref{figflu}).  The Benjamini--Hochberg
 corrected Fisher $p$-value is 0.017, and so the term would be considered enriched in the standard
 analysis;   it is also inferred to be active by MGSA (posterior probability 0.897). But it
 is not in the MAP estimate by MFA-ILP and its posterior activation probability is 0.000.
 Now 10 of these 12 influenza-involved GO:0032434 genes are part of the \textit{child} term
 GO:0032436, \textit{positive regulation of proteasomal ubiquitin-dependent protein catabolic process},
 a set of size 31 genes.  Both terms are
 found by the univariate Fisher procedure, exemplifying the redundancy issue; the more specific
 term GO:0032436 is identified by MFA-ILP and has high posterior activation probability.
 The MFA calculation favors the explanation whereby the smaller set is activated; this fails
 to cover two of the 12 GO:0032434 influenza genes, but it also simplifies
 the explanation of nonlisted genes in that set.  If the larger set (GO:0032434) is activated,
 then we have a lot more  mis-covered genes, that is, those in the set but  undetected by RNAi
 [$15 = (48-12) - (31-10)$].
 With this example, it may be that the more specific \textit{positive regulation}
 term better characterizes the experimental gene list.

 In Hao et al., gene set analysis was used to show that the four
 separate RNAi studies agreed more substantially than was evident by inspecting
 overlaps  among the four gene lists.
 It was applied separately to the study-specific
 gene lists, and then the agreement among these four set lists was measured. For
both Fisher's test and MGSA-ILP, the among-study set-level agreement was significant
according to a simple permutation calibration.  Curiously, the agreement by MGSA was
 not significant by that measure, owing primarily to the fact that
 very few sets were inferred to be active in the separate studies.
 The common set-level signal, in conjunction with other forms of meta-analysis, provided
 evidence that genome-wide RNAi studies have higher false-negative rates than false-positive
 rates.

\subsection{Other issues}

 The full effects of prior choice in model-based gene-set analysis require further investigation.
 As to the practical importance of one choice over another, we do not examine the biological
 distinctions between the inferences produced by different methods.  A close reading of the
 T2D and RNAi case studies above provides an initial indication of how and why reported set
 lists can differ, but assessing the biological significance of these differences is beyond
 the present scope.  The procedures have distinct statistical properties and
 MFA more efficiently captures the functional content of the reported gene
 list in terms of model fit.
  We point out that the distinctions present themselves when using the
 relatively complex GO system.  Control calculations  show that MGSA and MFA
 give essentially the same results when applied to the less complex KEGG system
 [Figures S1 and~S2, \citet{wa14}].

 In our comparisons we used MGSA
 to obtain an estimate of the hyper-parameter~$\pi$, which affects
 the overall rate of set activity.  In order to control the comparison,
 we used the same numerical value of $\pi$
 in the MFA calculations (in both MFA and MGSA we fixed the other  parameters $\alpha$ and $\gamma$ at
 externally derived values).   Further improvements of MFA may be possible using alternative
 estimates of $\pi$.  Other model elaborations which could be useful in some applications include
 extending MFA beyond binary $X_p$ and allowing dependence in gene level measurement errors.

 Compute times for MFA depend on the size and content of the gene list,  the incidence matrix $I$,
 and the model parameters:
 MFA-ILP used  2.5 CPU hours for T2D, and 23 CPU hours for RNAi;\footnote{R was
 running on a  4$\times$ AMD Opteron(TM) Processor 6174 (48 cores) with 128 GB RAM.} less time was required for MFA-MCMC (20 and 45 CPU minutes,
 resp.).

\begin{table}
\caption{Version effects:
 Tabulated are similarity scores comparing set structure and inferred active sets over
 time-adjacent versions of GO[5:50].  The collections of sets annotating between 5 and 50 human genes
 fluctuate over recent versions of GO. Respectively, in the four most recent fall versions of Bioconductor,
 the collections contain 4830, 5546, 6037 and 6488 gene sets.  The first row shows the Jaccard index (size of intersection over size
 of union) comparing subsequent versions of these gene-set collections.  In addition to the collections changing, the
 annotations recording which genes are in which sets also change over time.  The second row measures similarity of the
 sets of annotations.  Subsequent rows show similarity of reported lists of gene sets in the
two main examples. In this comparison, the set is reported if it is in the MAP estimate by MFA-ILP and if its marginal posterior
 probability exceeds a threshold (50\% or 80\%).  Inferred active sets depend to some extent on the GO version in view;
  setting stronger marginal posterior thresholds reduces the false-discovery rate and reduces the version effect} \label{tabversion}
\begin{tabular*}{\tablewidth}{@{\extracolsep{\fill}}lccc@{}}\hline
      &   \textbf{2010--2011} & \textbf{2011--2012} & \textbf{2012--2013} \\
      \hline
Sets  &      0.72   & 0.82  & 0.80 \\
Annotations & 0.54  & 0.70  & 0.66 \\
T2D  50/MAP &  0.33     & 0.60    & 0.30 \\
T2D  80/MAP &  1.00     & 0.75    & 0.60  \\
RNAi 50/MAP & 0.72  & 0.70  & 0.88 \\
RNAi 80/MAP & 0.90  & 0.79  & 0.85 \\
\hline
\end{tabular*}\vspace*{6pt}
\end{table}

 We have argued
 that temporal changes in GO reflect an increase in the complexity of the
 functional record that justifies the more refined prior distribution used
  in MFA  (Figure~\ref{figgo}).
These changes also
 tell us that the results of a given analysis naturally depend on the GO version,
 since the sets involved and the annotations of genes to sets continue to evolve. To assess the version effect,
 we applied MFA to four recent GO versions (2010--2013) in
 both the T2D and RNAi examples, and using sets annotating between 5 and 50 human genes (GO[5:50]).
 Table~\ref{tabversion} records how similar are time-adjacent versions of GO as well as how similar are
  results of MFA.
 The changes in GO reflect new terms and new annotations, as well as
 sets moving in and out of GO[5:50] as more genes become annotated.
 Against this substantial evolution of GO we see that
   MFA results do also change, but that the changes are less the more stringent is the marginal posterior
 probability cutoff.

 A feature of MGSA and MFA is that activation of the set implies activation of all genes in the set. This
 strict form of nonnull relationship is quite different from many univariate methods, which would claim a set is nonnull
 if any of its contained genes is nonnull\vadjust{\goodbreak} [e.g., the \textit{self-contained} tests of \citet{gb}].
 It is precisely this relationship, however, that enables multivariate (i.e., mult-set)  analysis, as the role model
 offers a straightforward approach to deal with the complex overlaps in the functional record.
 We note that the role model allows a weaker interpretation;
 for instance, we could continue to assert that a gene is activated if it is contained in any activated set, while
 allowing that only a fraction of activated genes are nonnull. The difference would be in the tabulation of errors
 (e.g., $X_p=0$ might not be a false negative when $A_p=1$) and in the interpretation of $\alpha$ and $\gamma$;
 the family of joint distributions would be the same. As GO and other repositories record functions of ever more
 specific gene combinations, it is reasonable to expect that a combination of genes relevant in the cells on
 test is within the repository. The strict interpretation of activation is parsimonious and is
 justified when the repository is sufficiently well endowed with relevant sets.  We performed a small simulation
 study, using the T2D data structure, to assess MFA's ability to recover small activated sets.  In each of 100
 simulated cases, we fixed the repository (GO[5:50]),  we randomized the response vector $X=(X_p)$ by appending
 to the T2D genes a randomly selected small ($5\mbox{--}10$ genes) set from GO, and we inferred the activated sets using MFA. In 91
 cases the appended set was identified as active by MFA-ILP, demonstrating in a limited way the ability of
 the methodology to recover signals represented in the repository.\vadjust{\goodbreak}

\section{Proofs}\label{sec7}

\subsection{Proof of Theorem~\texorpdfstring{\protect\ref{thm22}}{2.2}}

Relative to all the sets and parts in the system $I$, we say
AH holds if and only if
 $A_p= \max_{w\dvtx  p \in w} Z_w$ for all $p$ and $Z_w = \min_{p\dvtx  p \in w} A_p$
for all $w$.  Recall that all $A_p$ and $Z_w$ are binary, in $\{0,1\}$.
The first condition $A_p=\max_{w\dvtx  p \in w} Z_w$ implies
$A_p \geq Z_w$ for all $w$ with $p \in w$; that $A_p$ achieves the max of the
$Z_w$'s has to account for the possibility that $A_p=1$ when all $Z_w=0$, but
this is covered by having $A_p \leq \sum_{w\dvtx  p \in w} Z_w$.  Thus, the
condition $A_p = \max_{w\dvtx  p \in w} Z_w$ is equivalent to the first two constraints
in Theorem~\ref{thm22}.

To address the second condition, that $Z_w = \min_{p\dvtx  p \in w} A_p$ for all $w$,
 define a~new variable
\[
T_w = 1 + \sum_{p\dvtx  p \in w} ( A_p
- 1 ),
\]
and notice that $T_w=1$ if and only if $A_p=1$ for all $p \in w$,
 otherwise $T_w \leq 0$.  Observe that the second condition is equivalent to
\begin{equation}
\label{eqcondition}
\sum_{p\dvtx  p \in w} ( Z_w -
A_p + 1 ) - T_w \geq 0
\end{equation}
since if all $A_p=1$, for $p \in w$, then $T_w=1$, and $Z_w$ must equal 1 to
satisfy (\ref{eqcondition}); otherwise, if at least one of the $A_p$'s equals 0,
then $T_w \leq 0$, and noting that the summation in (\ref{eqcondition}) is
positive confirms the claim.  Next, replacing $T_w$ in (\ref{eqcondition}) with
its definition, we obtain the third stated inequality
\[
\sum_{p\dvtx  p \in w} ( Z_w - 2A_p +
2 ) \geq 1.
\]

\subsection{Proof of the Theorem~\texorpdfstring{\protect\ref{thm41}}{4.1}}
Compared to $Z_{w_0} = 0$, the possible maximal value added to the
objective function by letting $Z_{w_0}$ be 1 is  $c_1 + c_3\sum_{p\in P^+_{w_0}} 1$
(considering parts in $P_{w_0}^{-}$ may already be activated or $w_0$ has no parts in $P^-$,
the best case), however, which is negative since $w_0\in W^*$.
So $Z_{w_0}=0$ is preferred toward maximizing the objective function.
Next we need to prove that letting $Z_{w_0}=0$ and $A_{p_0}=0$ will not
violate the inequalities in Theorem~\ref{thm22}.

Denote by $W_0$ and $P_0$ the sets of $w_0$ and $p_0$ satisfying the state in the theorem,
respectively. We claim that for each $p_0\in P_0$, $\{w\dvtx  I_{p_0,w}=1\}\subset W_0$.
If not, then there exists $w^*\in W^*\setminus W_0$ such that $I_{p_0, w^*}=1$, so $w^*$
will be in $W_0$. This is a contradiction. Thus, $A_{p_0} = \max_{\{w\dvtx  I_{p_0,w}=1\}} Z_w = 0$,
so the first two AH inequalities are satisfied. It is readily verified
that the third inequality is also satisfied.

\section*{Acknowledgments}

  We thank Christina Kendziorski and anonymous reviewers for comments that helped to guide
this research.
 Both the supplementary material document and a software implementation of MFA are available
 at
 \surl{http://www.stat.wisc.edu/\textasciitilde newton/}. An initial version of this manuscript was released as
 Technical Report \#1174, Department of Statistics, University of Wisconsin, Madison. The current
 version is dated August, 2014.



\begin{supplement}[id=suppA]
\sname{Supplement}
\stitle{More on role modeling}
\slink[doi]{10.1214/14-AOAS777SUPP}  
\sdatatype{.pdf}
\sfilename{aoas777\_supp.pdf}
\sdescription{We provide further details on violation probabilities,
on estimating false-positive and true-positive error rates, on preparing
 data for the ILP algorithm, and on further data analysis findings in
 the T2D and RNAi examples.}
\end{supplement}



\printaddresses
\end{document}